\documentclass[11pt,twoside]{article}

\usepackage{natbib}
\usepackage{asp2006}
\usepackage{graphicx}

\def\ltsima{$\; \buildrel < \over \sim \;$}
\def\simlt{\lower.5ex\hbox{\ltsima}}
\def\gtsima{$\; \buildrel > \over \sim \;$}
\def\simgt{\lower.5ex\hbox{\gtsima}}

\newcommand\mearth{{\,{\rm M}_{\oplus}}}

\begin{document}
\title{A new view on planet formation}   
\author{Sergei Nayakshin}
\affil{Department of Physics \& Astronomy,
  University of Leicester, Leicester, LE1 7RH, UK}

\begin{abstract} 
The standard picture of planet formation posits that giant gas planets are
over-grown rocky planets massive enough to attract enormous gas
atmospheres. It has been shown recently that the opposite point of view is
physically plausible: the rocky terrestrial planets are former giant planet
embryos dried of their gas ``to the bone'' by the influences of the parent
star. Here we provide a brief overview of this ``Tidal Downsizing'' hypothesis
in the context of the Solar System structure.
\end{abstract}



\section{Introduction}

In the popular ``core accretion'' scenario \citep[CA model hereafter;
  e.g.,][]{Safronov69,Wetherill90,PollackEtal96}, the terrestrial planet cores
form first from much smaller solid constituents. A massive gas atmosphere
builds up around the rocky core if it reaches a critical mass of about $10
\mearth$ \citep[e.g.,][]{Mizuno80}. The CA model's main theoretical difficulty
is in the very beginning of the growth: it is not clear how metre-sized rocks
would stick together while colliding at high speeds, subject to high radial
drifts into the parent star \citep{Weiden77,Weiden80}, although gas-dust
dynamical instabilities are suggested to help 
\citep[e.g.,][]{YoudinGoodman05,JohansenEtal07}. Nevertheless, believed to be
the only viable model for terrestrial planet formation, the model has enjoyed
an almost universal support \citep[e.g.,][]{IdaLin08}.

This strongest asset of the theory -- a ``monopoly'' on making terrestrial
planets -- is actually void. Recently, it has been proposed by
\cite{BoleyEtal10,Nayakshin10a,Nayakshin10b,Nayakshin10c} that a modified
version of the gravitational disc instability model for giant planet
formation\citep{Kuiper51,Boss98} may account for terrestrial planets as well,
if gas clump migration \citep{GoldreichTremaine80} and clump disruption due to
tidal forces \citep{McCreaWilliams65} are taken into account. This new scheme
addresses \citep{Nayakshin10c} all of the well known objections
\citep{Wetherill90,Rafikov05} to forming Jupiter in the Solar System via disc
fragmentation.

The TD hypothesis is a new combination of earlier ideas and contains four
important stages (Figure 1):
\begin{itemize}
\item[(1)] Formation of gas clumps (which we also call giant planet embryos;
  GEs). As the protoplanetary disc cannot fragment inside $R \sim 50$
  AU \citep{Rafikov05,BoleyEtal06}, GEs are formed at somewhat larger radii. The
  mass of the clumps is estimated at $M_{\rm GE} \sim 10 M_J$ (10 Jupiter
  masses) \citep{BoleyEtal10,Nayakshin10a}; they are intially fluffy and cool
  ($T\sim 100$ K), but contract with time and become much
  hotter \citep{Nayakshin10a}.

\item[(2)] Inward radial migration of the clumps due to gravitational
  interactions with the surrounding gas
  disc \citep{GoldreichTremaine80,VB10,BoleyEtal10,ChaNayakshin10}.

\item[(3)] Grain growth and sedimentation inside the
  clumps \citep{McCreaWilliams65,Boss98,BossEtal02}. If the clump temperature
  remains below $1400-2000 $K, massive terrestrial planet cores may
  form \citep{Nayakshin10b}, with masses up to the total high Z element content
  of the clump (e.g., $\sim 60$ Earth masses for a Solar metalicity clump of $10
  M_J$).

\item[(4)] A disruption of GEs in the inner few AU due to tidal
  forces \citep{McCrea60,McCreaWilliams65,BoleyEtal10,Nayakshin10c} or due to irradiation
  from the star \citep{Nayakshin10c} can result in (a) a smallish solid core and
  a complete gas envelope removal -- a terrestrial planet; (b) a massive solid
  core, with most of the gas removed -- a Uranus-like planet; (c) a partial
  envelope removal leaves a gas giant planet like Jupiter or Saturn.  For (b),
  an internal energy release due to a massive core formation removes the
  envelope \citep{HW75,Nayakshin10b}.
\end{itemize}

It is interesting to note that it is the proper placement of step (1) into the
outer reaches of the System and then the introduction of the radial migration
(step 2) that makes this model physically viable.  The theory based on
elements (3,4) from an earlier 1960-ies scenario for terrestrial planet
formation by \cite{McCrea60,McCreaWilliams65} were rejected by \cite{DW75} because step
(1) is not possible in the inner Solar System. Similarly, the giant disc
instability \citep{Kuiper51,Boss98} cannot operate at $R\sim 5$ AU to make
Jupiter \citep{Rafikov05}.  It is therefore the proper placement of step (1)
into the outer reaches of the System and then the introduction of the radial
migration (step 2) that makes this model physically viable. The new hypothesis
resolves \citep{Nayakshin10d} an old mystery of the Solar System: the mainly
coherent and prograde rotation of planets, which is unexpected if planets are
built by randomly oriented impacts.

\section{Solar System structure}

The gross structure of the Solar System planets is naturally accounted for by
the TD model. The innermost terrestrial planets are located within the tidal
disruption radius of $r_t \sim 2-3$ AU \citep{Nayakshin10c}, so these are
indeed expected to have no massive atmospheres. The asteroid belt in this
scheme are the solids that grew inside the giant planet embryos but not made
into the central core, and which were then left around the $r_t$. The gas
giant planets are somewhat outside the tidal disruption radius, and thus have
been only partially affected by tidal disruption/Solar irradiation.

The outer icy giant planets are too far from the Sun to have been affected
strongly by it, so they are interesting cases of {\em self-disruption} in the
TD model. In particular, 35 years ago, \cite{HW75} suggested that the massive
core formation in Uranus and Neptune evaporated most of their hydrogen
envelopes.  To appreciate the argument, compare the binding energy of the
solid core with that of the GE.  We expect the core of high-Z elements to have
a density $\rho_c \sim $ a few g cm$^{-3}$. The radial size of the solid core,
$R_{\rm core} \sim (3M_{\rm core}/4\pi\rho_c)^{1/3}$. The binding energy of
the solid core is
\begin{equation}
E_{\rm bind, c} \sim \frac{3}{5} \frac{G M_{\rm core}^2}{R_{\rm core}} \approx
10^{41} \; \hbox{erg}\; \left(\frac{M_{\rm c}}{10 \mearth}\right)^{5/3}\;.
\label{ebind_p}
\end{equation}
The clump radius $R_{\rm GE} \approx 0.8$ AU at the age of $t=10^4$ years,
independently of its mass\cite{Nayakshin10c}, $M_{\rm GE}$.  Thus, the GE
binding energy at that age is
\begin{equation}
E_{\rm bind, GE} \sim \frac{3}{10} \frac{G M_{\rm GE}^2}{R_{\rm GE}}
\approx 10^{41} \; \hbox{erg}\; \left(\frac{M_{\rm GE}}{3 M_J}\right)^{2}\;.
\label{ebind_p}
\end{equation}
The two are comparable for $M_{\rm core} \sim 10 \mearth$.  Radiation
hydrodynamics simulations confirm such internal disruption events: the run
labelled M$0\alpha3$ in \cite{Nayakshin10b} made a $\sim 20 \mearth$ solid
core that unbound all but $0.03 \mearth$ of the gaseous material of the
original $10 M_J$ gas clump.

Future work on the TD hypothesis should address the outer Solar System
structure (Kuiper belt; comet compositions, etc.). Detailed predictions for
exo-planet observations are difficult as the model dependencies are non-linear
\citep{Nayakshin10b}, but some predictions distinctively different from the CA
scenario may be possible as planets loose rather than gain mass as they
migrate inwards.


\begin{figure}[!ht]
\begin{center}
\includegraphics[scale = 0.5]{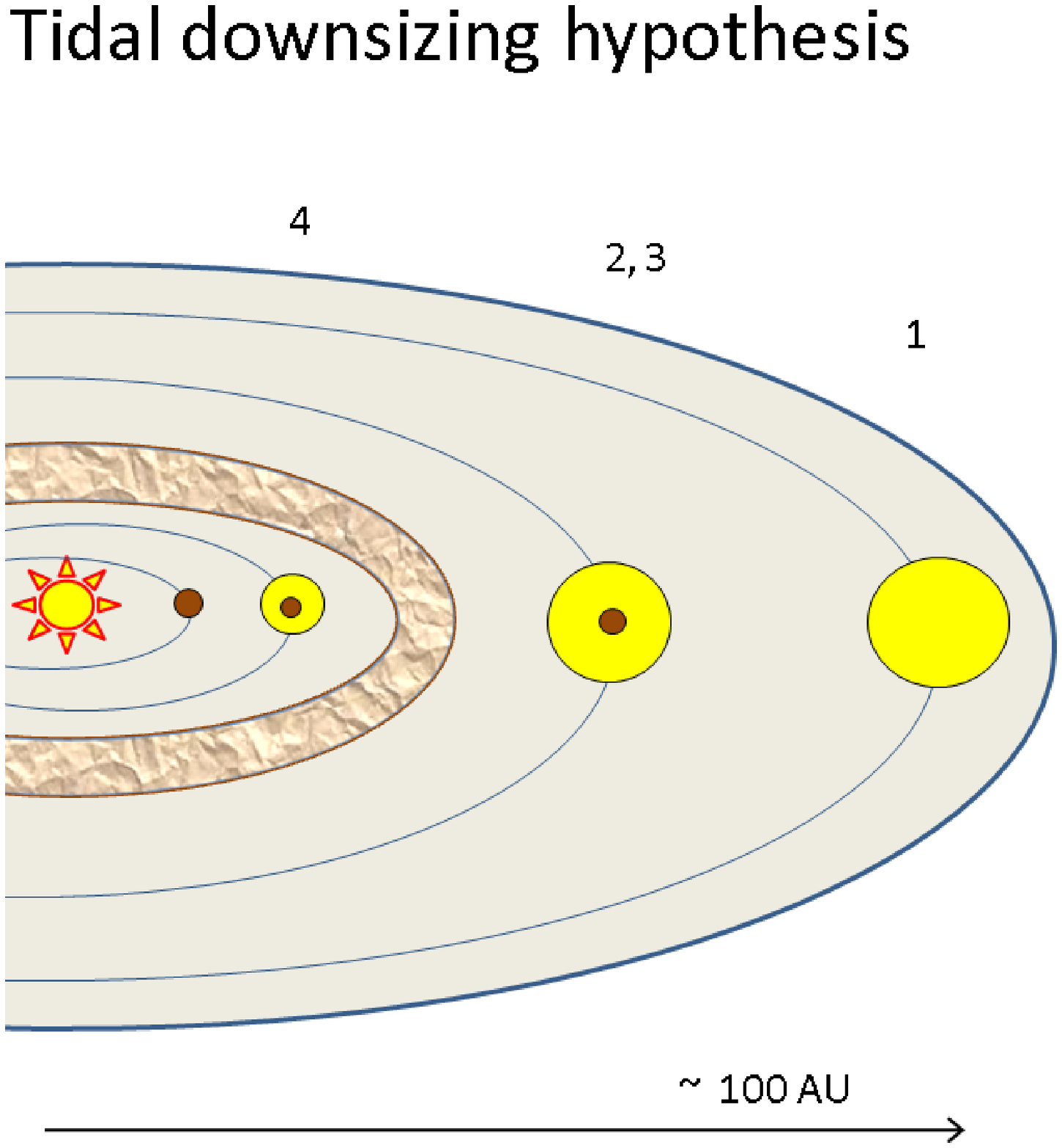}
\end{center}
\caption{A cartoon of the Tidal Downsizing hypothesis. A protostar (the
  central Sun symbol) is surrounded by a massive $R\simgt 100$ gas disc (the
  larger grey oval). The four planet formation stages are schematically marked
  by numbers: (1) The formation of massive gas clumps (embryos) in the outer
  disc; (2) migration of the clumps closer in to the star, occurring
  simultaneously with (3) dust grains growth and (possibly) sedimentation into
  a massive solid core in the centre. The core is shown as a small brown
  sphere inside the larger gas embryo; (4) disruption of the embryo by tidal
  forces, irradiation or internal heat liberation. The brown pattern-filled
  donut-shaped area shows the solid debris ring left from an embryo
  disruption.  The
  most inward orbit in the diagram shows a terrestrial-like planet, e.g., a
  solitary solid core whose gas envelope was completely removed. The planet on
  the next smallest orbit is a giant-like planet with a solid core that
  retained some of its gas envelope.}
\end{figure}

\acknowledgements The author acknowledges the support of the STFC research
council and the IAU travel grant to attend this exciting meeting. 




\end{document}